\documentclass[aps,prl,superscriptaddress,longbibliography,twocolumn]{revtex4-1}

\usepackage{graphicx}
\usepackage{epstopdf}
\usepackage{amssymb}
\usepackage{color}
\usepackage{enumerate}
\usepackage{ulem}
\usepackage{setspace}

\begin{document}

\title{Collective energy gap of preformed Cooper-pairs in disordered superconductors}

\author{Thomas Dubouchet}
\affiliation{Univ. Grenoble Alpes, CEA, INAC, PHELIQS, F-38000 Grenoble, France}
\author{Benjamin Sac\'{e}p\'{e}}
\email[E-mail: ]{benjamin.sacepe@neel.cnrs.fr}
\affiliation{Univ. Grenoble Alpes, CNRS, Grenoble INP, Institut N\'{e}el, 38000 Grenoble, France}
\author{Johanna Seidemann}
\affiliation{Univ. Grenoble Alpes, CNRS, Grenoble INP, Institut N\'{e}el, 38000 Grenoble, France}
\author{Dan Shahar}
\affiliation{Department of Condensed Matter Physics, Weizmann Institute of Science, Rehovot 76100, Israel}
\author{Marc Sanquer}
\affiliation{Univ. Grenoble Alpes, CEA, INAC, PHELIQS, F-38000 Grenoble, France}
\author{Claude Chapelier}
\email[E-mail: ]{claude.chapelier@cea.fr}
\affiliation{Univ. Grenoble Alpes, CEA, INAC, PHELIQS, F-38000 Grenoble, France}

\date{\today}

\begin{abstract}
\textbf{
In most superconductors the transition to the superconducting state is driven by the binding of electrons into Cooper-pairs. The condensation of these pairs into a single, phase coherent, quantum state takes place concomitantly with their formation at the transition temperature, $T_c$. A different scenario occurs in some disordered, amorphous, superconductors: Instead of a pairing-driven transition, incoherent Cooper pairs first pre-form above $T_c$, causing the opening of a pseudogap, and then,  at $T_c$, condense into the phase coherent superconducting state. Such a two-step scenario implies the existence of a new energy scale, $\Delta_{c}$, driving the collective superconducting transition of the preformed pairs. Here we unveil this energy scale by means of Andreev spectroscopy in superconducting thin films of amorphous indium oxide. We observe two Andreev conductance peaks at $\pm \Delta_{c}$ that develop only below $T_c$ and for highly disordered films on the verge of the transition to insulator. Our findings demonstrate that amorphous superconducting films provide prototypical disordered quantum systems to explore the collective superfluid transition of preformed Cooper-pairs pairs.}
\end{abstract}

\maketitle

While a small amount of (non-magnetic) disorder weakly affects superconductivity~\cite{Anderson59, Abrikosov59},  the situation changes radically when disorder reaches a level such that localization of electronic wave functions develops. In some amorphous materials, such an extreme situation in which the normal state electron transport is no longer metallic due to the abundant scattering, yields a fragile superconducting state at low temperature. On increasing disorder further or applying a magnetic field this superconducting state undergoes a sharp quantum phase transition to insulator~\cite{Goldman98, Gantmakher10}.

Recently a body of tunneling experiments performed on different disordered thin films~\cite{Sacepe08,Sacepe10, Mondal11,Sacepe11,Chand12,Sherman12, Noat13,Ganguly17} have shown that the spectral properties of superconductivity bordering the transition to insulator deviate significantly from the Bardeen-Cooper-Schrieffer (BCS) theory~\cite{BCS}. The new picture drawn by these measurements is that electrons pre-form into Cooper-pairs~\cite{Sacepe11} well above the transition temperature, $T_c$, leading to a deep suppression of the single-particle density-of-states (DOS) at the Fermi level, which is known as the pseudogap anomaly~\cite{Sacepe10,Mondal11,Sacepe11}. Only at a temperature $T=T_c$ at which resistivity vanishes and global phase coherence between these preformed pairs sets in, the pseudogap evolves into a hard gap, $E_{g}$. The systematic observation of an anomalously large and spatially fluctuating spectral gap $E_g$, with a ratio $E_{g}/k_BT_c$ reaching up to $\sim 6$ in the most disordered samples, has pointed out a dissociation between $T_c$ and $E_{g}$, in stark contrast to BCS superconductors for which $E_{g}/k_BT_c=1.76$ ($k_B$ is the Boltzmann constant). 

On these grounds theory predicts that a second energy scale, $\Delta_{c}$, may be at play to drive the collective, BCS-type, superconducting transition of the preformed pairs~\cite{Feigelman07,Feigelman10}. This collective gap is expected to scale with $T_c$ and thus lies within the single-particle gap $E_{g}$, making tunneling measurement unable to unveil it.

As earlier proposed in the context of high $T_c$ superconductors~\cite{Deutscher99,Deutscher05}, the means to unveil $\Delta_{c}$ consists of coherent transfer of pairs of electrons from a normal metal (N) electrode to the superconductor  (S). The mechanism that allows charge transfer through the N/S interface correlates, in the metal, an impinging electron with another one, having energy below the Fermi level, to form a Cooper-pair in the superconductor. This so-called Andreev process results from the particle-hole mixing occurring in the superconductor within the energy interval of the BCS superconducting gap~\cite{Andreev64}. On the other hand, in a superconductor with preformed Cooper-pairs, theory predicts that the energy window for particle-hole mixing no longer coincides with the single particle gap $E_g$ but instead with $\Delta_{c}$~\cite{Feigelman10}. Consequently, whereas single-particle tunneling probes only $E_g$, two-particles spectroscopy, termed Andreev spectroscopy, enables direct measurement of the collective energy gap $\Delta_{c}$. 

Here we report on Andreev spectroscopy performed on superconducting thin films of amorphous indium oxide (a:InO) on the verge of the insulating state~\cite{Shahar92}. The transfer of electron pairs being a second order process as compared to single electron tunneling, Andreev spectroscopy requires high contact transparency. We therefore used the metallic tip of a scanning tunneling microscope (STM) to control the transparency, that is, the conductance, of the point-contact formed between the tip and the film, and perform a systematic study from single-particle tunneling to Andreev spectroscopy~\cite{Agrait92}.

\begin{figure}[h!]
\includegraphics[width=1\columnwidth, bb = 0 20 600 480] {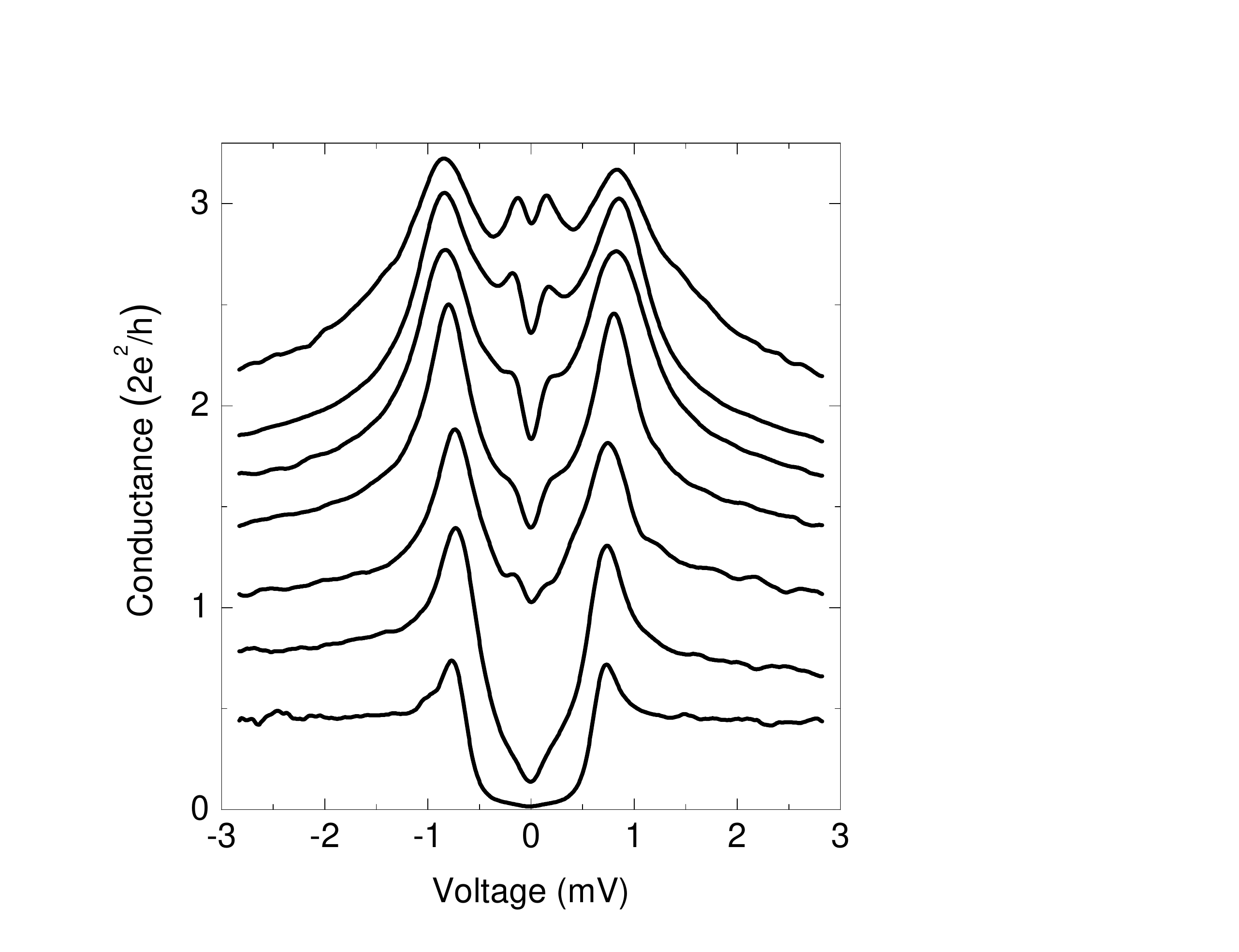}
\caption{\textbf{From tunneling to Andreev spectroscopy.} Evolution of the local differential conductance $G=dI/dV$ measured at $T=0.065\,$K versus voltage and at different values of the point-contact conductance $ G(V=-3$ mV) for  sample InO-2. Conductance is normalized to $2e^2/h$.}  
\label{figure1}
\end{figure}

The chief result of this work is presented in figure \ref{figure1} where we display a set of differential conductance curves $G(V)=dI/dV$ versus bias-voltage
 $V$ across a point-contact on sample InO-2. For each curve the point-contact conductance is controlled by adjusting the current set-point of the STM feedback loop for a given voltage-bias ($V=-3$ mV). From bottom to top curve the point-contact conductance varies from $e^2/h$ to more than $4e^2/h$ ($e$ is the electron charge; $h$ is the Plank constant). 

We begin by noting that for the low point-contact conductance (botton curve in Fig~\ref{figure1}) $G(V)$ resembles that of single-particle tunneling with a gap $E_g$ at the Fermi level bordered by single-particle peaks ( see Fig.~S2 in SI). Increasing further the conductance profoundly modifies the spectrum. While the single-particle peaks remains, the gap fills in with a significant level of conductance indicating the opening of new conduction channels for two-particles as the transparency rises. The Andreev spectroscopy regime is hence clearly established. 
\begin{figure}[h!]
\includegraphics[width=1\linewidth, bb = 0 120 780 480]{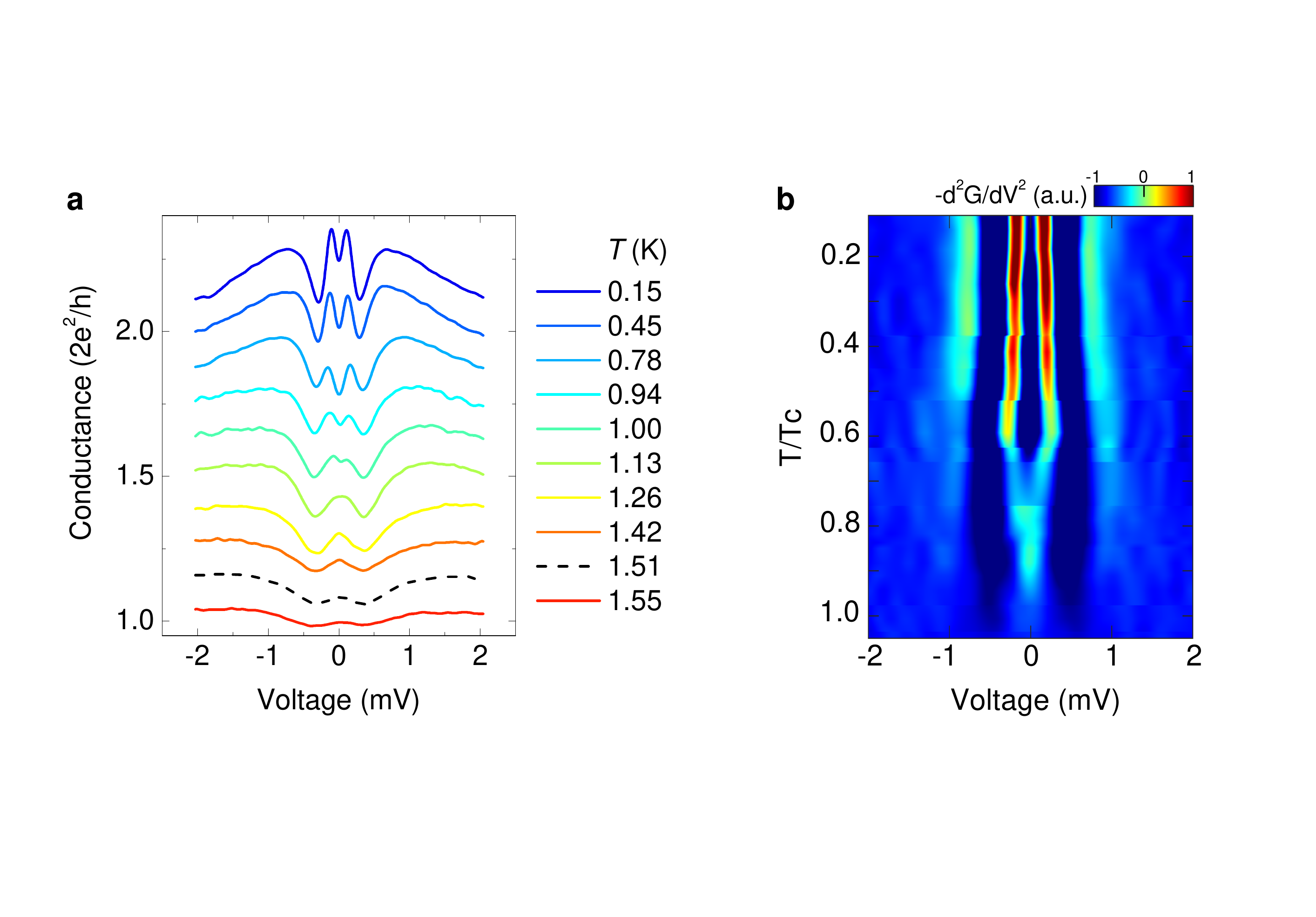}
\caption{{\bf Temperature dependence of the  Andreev spectroscopy.} \textbf{a,} Differential conductance curves in unit of $2e^2/h$ measured versus bias voltage at different temperatures. All curves except the bottom one are vertically shifted for clarity. The black dashed line corresponds to the spectrum measured at $T_c$. \textbf{b,} Colormap of $-d^2G/dV^2$ (arbitrary unit) versus bias voltage and reduced temperature $T/T_c$ for the whole set of data.}  
\label{figure2}
\end{figure}

For conventional, low disorder superconductors, Andreev process at the interface leads to a monotonous increase of conductance inside the gap ultimately reaching, for perfect transparency, twice the contact conductance~\cite{BTK82}. Our results on a highly disordered film displayed in figure \ref{figure1} are rather different. Two peaks inside the single-particle gap emerge out of the Andreev conductance background as the contact conductance rises above $2e^2/h$. Their energies are symmetric with respect to zero-bias, at $\pm 150 \, \mu V$, and independent of the contact transparency. These new peaks unveiled by Andreev spectroscopy are the main focus of this work and provide, as demonstrated below, compelling evidence for the collective gap $\Delta_{c}$ for the preformed Cooper pairs.

To trace back the superconductivity-related origin of these peaks we studied the $T$-evolution of the Andreev spectra measured at different point-contact locations of the samples.  One representative $T$-evolution of $G(V)$ measured on sample InO-2 is shown in Figure \ref{figure2}a. The point contact transparency is set large enough such that both Andreev and single-particle peaks are apparent. On increasing temperature the two Andreev peaks move towards zero-bias voltage at $T \gtrsim 0.6\,T_c$ and merges farther into a conductance maximum at zero-bias. This maximum, still just visible at $T=T_c$, disappears beyond $T \simeq 1.05\,T_c$. The $T$-evolution of the Andreev peaks is even more clearly seen in Fig. \ref{figure2}b by plotting $-d^2G/dV^2$ versus $T$ and $V$ (see Fig.~S3 in SI). 

The termination of the Andreev peaks at $T \simeq T_c$ that we observe in all our samples clearly indicates that they relate to coherent superconductivity.  This is in stark contrast with the $T$-evolution of the single-particle tunneling gap $E_g$, which remains unchanged at $T_c$, and only vanishes further at several times $T_c$, resulting in the appearance of the pseudogap in the single-particle density-of-states~\cite{Sacepe11}. 

On a theoretical standpoint a body of works on disordered superconductivity now converges to account for the pseudogap by the pre-formation of Cooper-pairs above $T_c$~\cite{Feigelman07,Feigelman10,Bouadim11}. The mechanism behind this relies on the delicate interplay between quantum localization of the single-particle states and the attractive pairing. As a result the anomalously large single-particle gap $E_g$ is predicted to embody two contributions. The first is the pairing energy gap $\Delta_p$ for the pre-formation of Cooper-pairs --the energy gain to pair electrons or equivalently the energy cost to add an odd number of electrons in localized orbitals-- at the origin of the pseudogap. The second  is the collective energy gap $\Delta_{c}$ related to the BCS-type condensation of the preformed pairs into the superconducting state~\cite{Feigelman10}. As $\Delta_p$ is not involved when transferring two electrons simultaneously, $\Delta_{c}$ only can give rise to the Andreev conductance signal that we observe in our data. 

\begin{figure}[h!]
\includegraphics[width=1\columnwidth, bb = 0 200 830 550 ]{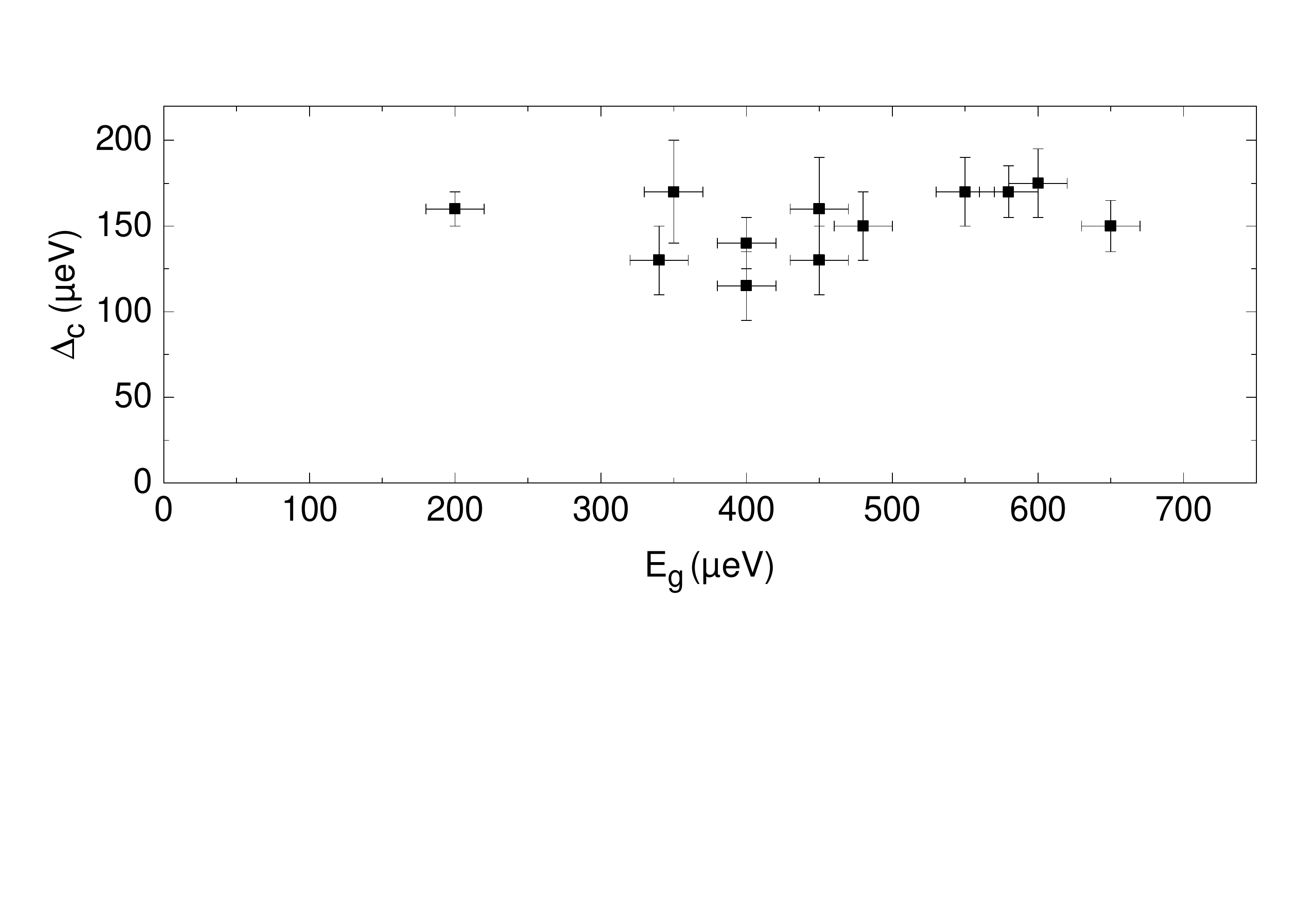}
\caption{\textbf {Collective gap versus spectral gap.} Collective energy gap $\Delta_c$ as a function of single particle gap $E_g$ measured at low temperature in contact and tunneling regime respectively. The data were obtained at different locations on three highly disordered a:InO samples. The error bars indicate the experimental accuracy in determining $\Delta_c$ and $E_g$.}
\label{figure3}
\end{figure}
Furthermore, we  investigate how the values of $E_g$ and $\Delta_{c}$ spreads over three highly disordered samples (samples InO-1, InO-2 and InO-3). All our measurements of $E_g$ and $\Delta_{c}$ performed on these samples are summarized in Figure~\ref{figure3}. The value of the local gap $E_g$ obtained by fitting the density of states of spectra in the tunneling regime (see Fig. S2 in SI) is strongly tip-position dependent with an an energy varying by a factor of more than three, that is, spreading between $200 \mu eV $ and $650 \mu eV $, as previously reported~\cite{Sacepe11}. In contrast, $\Delta_c$ scatters around an average value of $145\,\mu$eV with a variability of $\pm 30\, \mu$eV  that stems mainly from the experimental accuracy. Notice that this observation rules out multiple Andreev reflections at sub-multiple energies $2E_g/n$, $n$ being an integer, as a possible origin for the observed sub-gap structure in our spectra~\cite{Klapwijk82, Octavio83}.  Importantly, this average value of $\Delta_c$ that translates to $1.7 \pm 0.3\,$K is close to the $T_c$'s of the samples ($1.45$, $1.5$ and $1.9\,$K, see Table S1 in SI).  

\begin{figure}[h!]
\includegraphics[width=1\columnwidth, bb = 0 50 500 380 ]{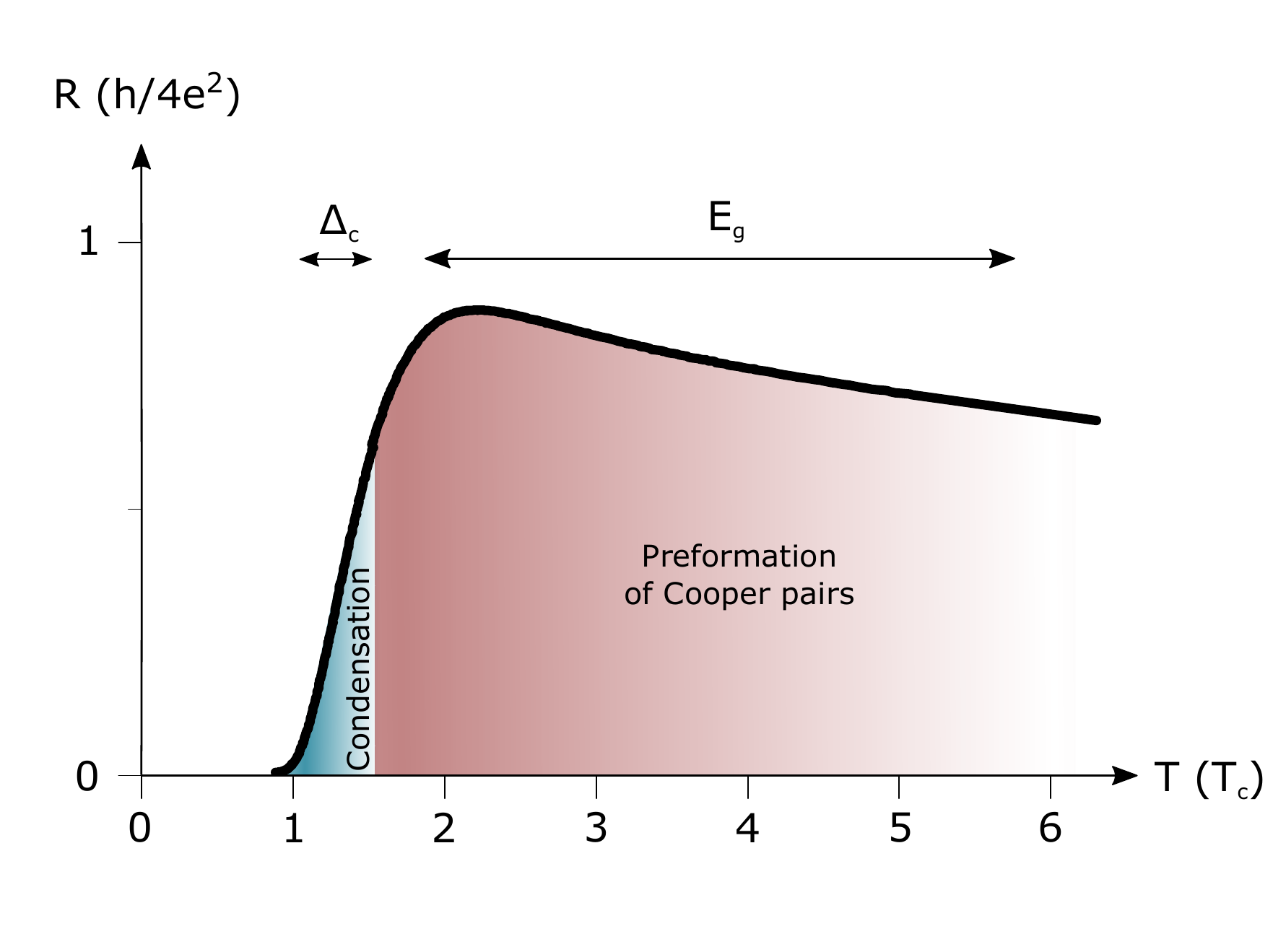}
\caption{{\bf Two-step transition to superconductor.} Resistance in unit of $h/4e^2$ versus $T$ normalized to $T_c$ illustrating the typical superconducting transition of a nearly critical thin films in the vicinity of the superconductor-insulator transition. The typical distribution of $\Delta_c$ and $E_g$ are indicated by the black arrows.}
\label{figure4}
\end{figure}

The two energy scales discussed in this work span on two clearly distinct ranges that are illustrated in Figure~\ref{figure4} where we have reported the respective energy ranges for $\Delta_{c}$ and $E_g$ on a typical resistance curve around the transition to the superconducting state. The condensation energy $\Delta_{c}$ is of the order of the critical temperature $T_c$, in stark contrast with the pseudogap $E_g$, which extends up to $\sim6T_c$. This coincidence between $\Delta_{c}$ and $T_c$ together with the suppression of $\Delta_{c}$ in the close vicinity of $T_c$ provide strong evidence that this new energy scale relates to the set in of macroscopic, superconducting phase coherence. 


Finally, to demonstrate that the Andreev peaks stem from the high level of disorder in our films and the proximity to the insulator, we conducted identical measurements on a low disorder a:InO film (sample InO-4) that behaves in all points as a conventional dirty superconductor~\cite{Sacepe15}, that is, without a pseudogap above $T_c$ (see Fig.~S4 in SI). The corresponding Andreev spectroscopy shows a standard, monotonous increase of the conductance within the single-particle gap, without any intra-gap peak (see Fig.~S5 in SI). This evolution, which has been observed for any position of the STM tip can be accurately described by Blonder-Tinkham-Klapwijk (BTK) theory for N/S  interfaces~\cite{BTK82}.

To conclude, the scenario drawn by our observations contrasts with standard BCS superconductors. The second energy scale we uncover together with the pseudogap comply with a two-step transition to superconductor, which develops, first, far above $T_c$ with the preformation of Cooper-pairs and follows at lower $T$ by the collective condensation of the preformed pairs into a macroscopically phase coherent superconducting state. The nature of such a superconducting state that develops in disordered superconductors bordering insulators, defies conventional theories and shall inspire new theoretical paradigms for superconductivity.

\section{Methods}

\textbf{Samples} Our samples are disordered thin films of amorphous indium oxide. The films are prepared by electron-beam evaporation of high purity ($99.99\%$) In$_{2}$O$_{3}$ onto SiO$_{2}$ in an O$_{2}$ partial pressure. Samples were patterned into Hall bridges via a shadow mask to perform \textit{in-situ} four-terminal transport measurements in the STM set-up and thus direct comparison between macroscopic and microscopic superconducting properties.\\

\textbf{Measurements} Tunneling and Andreev spectroscopy measurements were performed with a home-built STM cooled down to $0.05\,$K in an inverted dilution refrigerator. By varying the STM setpoint current from 1 nA to 1 $\mu$A with a bias voltage of a few millivolts, we tuned the junction resistance in the range $1-1000\,k\Omega$ and thus explored continuously the whole transition from tunneling to contact regime. The differential conductance $G(V)=\frac{dI}{dV}$ of the junction was measured with a lock-in amplifier technique with a voltage modulation of $10-30 \, \mu \text{V}$.\\

\textbf{Acknowledgments} We thank M. Feigel'man, L. Ioffe and Y. Nazarov for fruitful discussions. This research was supported in part by the French National Agency ANR-10-BLANC-04030-POSTIT, ANR-16-CE30-0019-ELODIS2 and the H2020 ERC grant \textit{QUEST} No. 637815.\\


\clearpage
\setcounter{figure}{0}
\setcounter{section}{0}
\renewcommand{\thefigure}{S\arabic{figure}}

\begin{center}
\section{Supplementary Information}
\end{center}

\section{I Transport measurements}
In this section we present the transport properties of the four amorphous InO films studied in this work. Their respective thickness, maximum sheet resistance above the transition and superconducting critical temperature are summarized in table \ref{table1}. For a given thickness of the film (InO-1,InO-2 and InO-3), the level of disorder is tuned by the partial oxygen pressure $P_{O_2}$ used during the In$_2$O$_3$ evaporation.

\begin{table}[h!]
\centering
\begin{tabular}{|c|c|c|c|c|}
\hline
\centering
Label & $T_c$ \textrm{(K)} & $R_{\square}^{max} (\Omega) $ & $d$ (\AA)  \\ \hline\hline
InO-1 & 1.45 & 5770 & 300 \\ \hline 
InO-2 & 1.5 & 5615 & 300 \\ \hline
InO-3 & 1.9 & 3885 & 300  \\ \hline
InO-4 & 3.2 & 400 & 700  \\ \hline

\end{tabular}
\caption{{\bf Sample characteristics.}\label{table1} Critical temperature $T_c$, maximum sheet resistance $R_{\square}^{max}$  and sample thickness $d$ of the four samples presented in Figure~\ref{figureS1}.}
\end{table}

\begin{figure}[h!]
\includegraphics[width=0.9\linewidth, bb = 50 50 750 450 ]{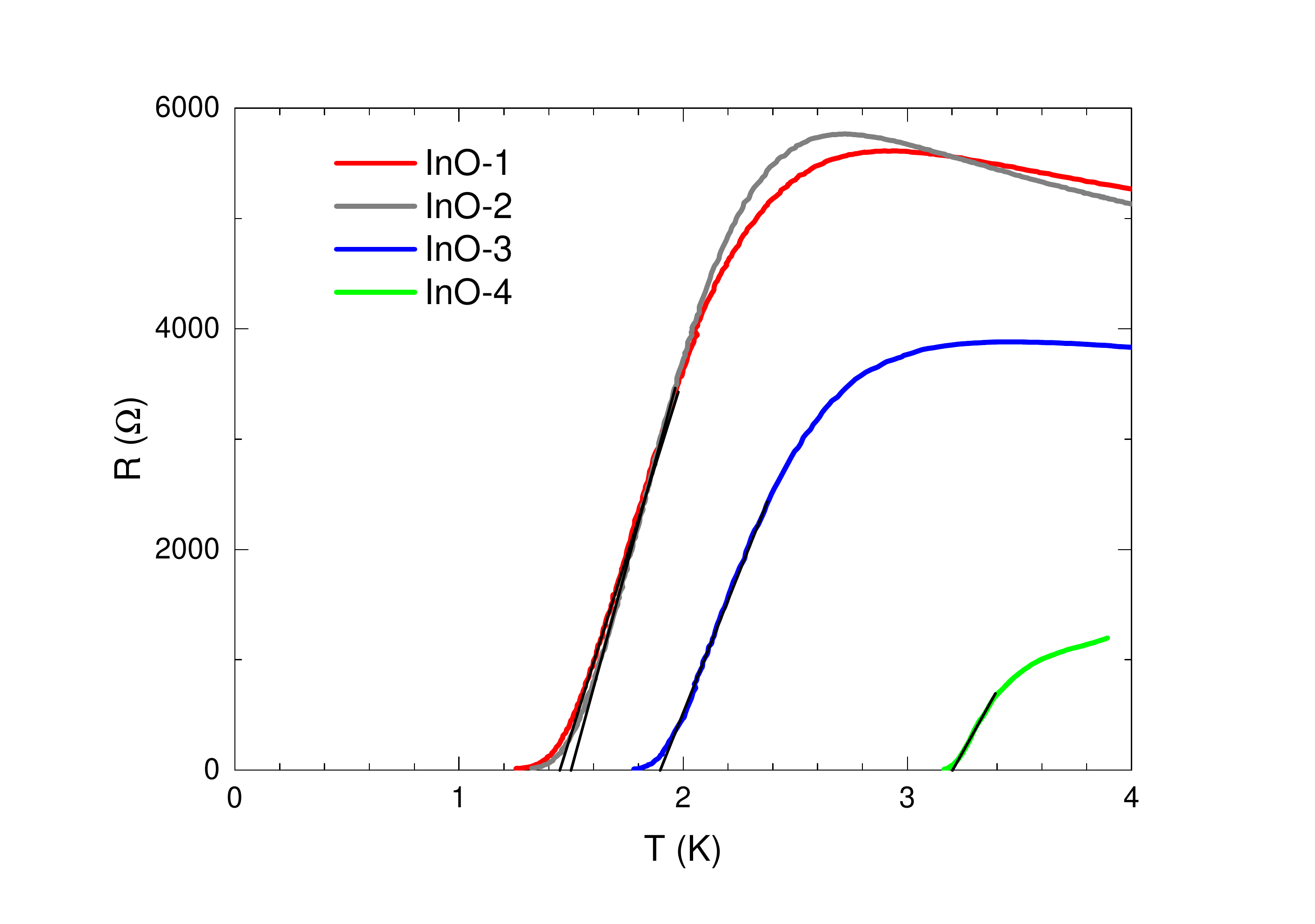}
\caption{{\bf Superconducting transition of the indium oxide films.} T-evolutions of the resistance obtained during the cooling of the STM set-up. The black solid lines are linear extrapolations of the superconducting drop.}
\label{figureS1}
\end{figure}

The $T$-evolutions of the resistivity for the samples are shown in Figure~\ref{figureS1}. The critical temperature, $T_c$, is defined as the extrapolation to R=0 of the resistance drop. $T_c$ is in the range $1-2\,K$ for the three highly disordered films, InO-1, InO-2 and InO-3. This places them at the verge of the superconductor-insulator transition~\cite{Shahar92,Gantmakher96}. The fourth film, InO-4, is much less disordered with $T_c$ above $3\,K$ and behaves as a usual BCS superconductor~\cite{Sacepe15}. 

\section{II Extraction of the spectral gap from tunneling spectra}

In order to estimate the spectral gap $E_{g}$ of the one-particle density of states, we fit the low conductance point-contact data with BCS theory. Figure S2 shows a typical tunneling spectrum measured for a tunneling conductance of $0.84 \,e^2/h$, and a BCS fit (red curve) with $E_{g} = 650 \, \mu eV$ and a Dynes parameter $\Gamma = 40 \, \mu eV$~\cite{Dynes86}.

\begin{figure}[h!]
\includegraphics[width=0.9\linewidth, bb = 50 50 750 500 ]{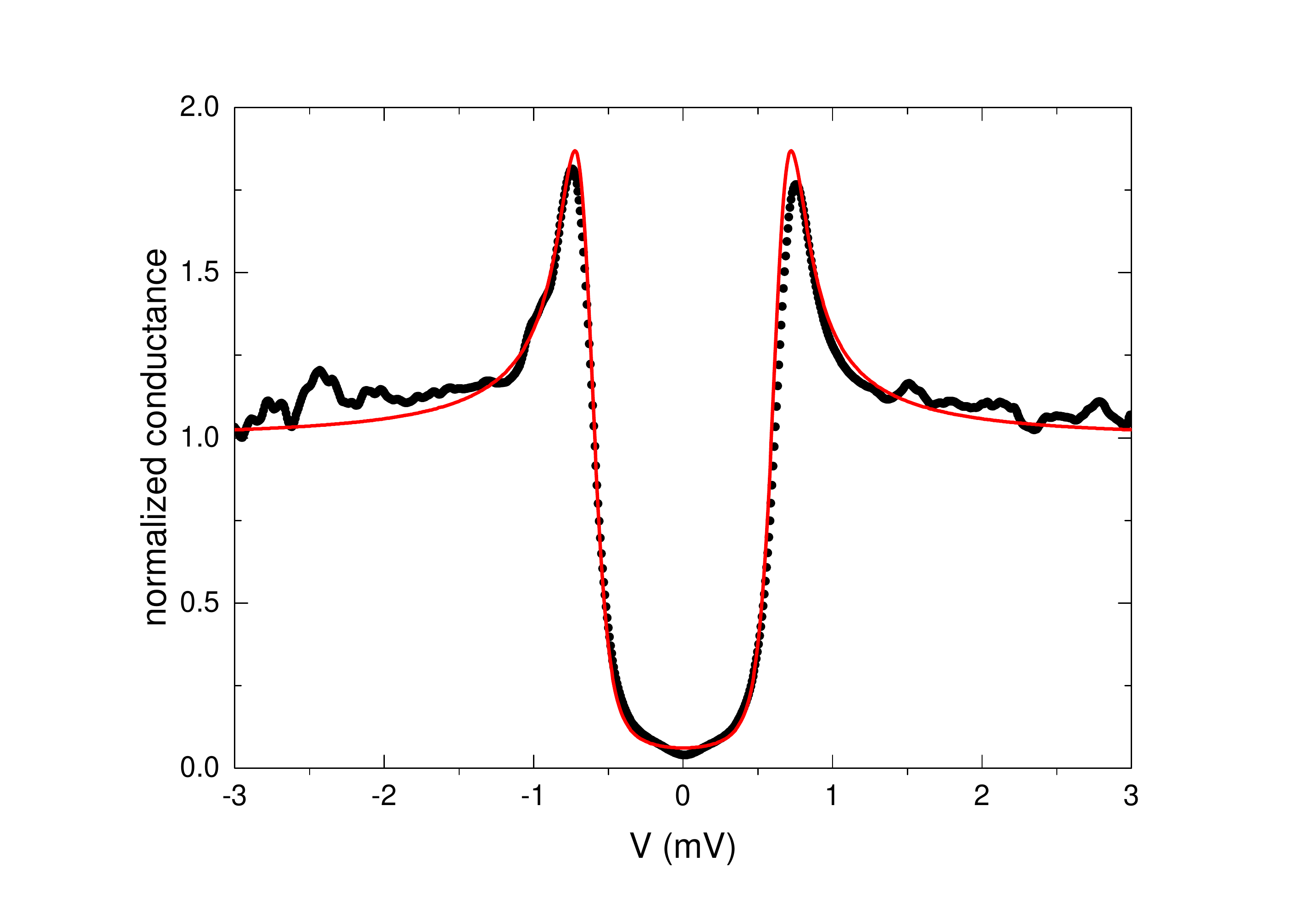}
\caption{{\bf Tunneling density of states.} Spectrum measured with a tunneling conductance of $0.84 \,e^2/h$. The black dotted curve represents normalized experimental data. The red curve is a BCS fit.}
\label{figureS2}
\end{figure}

\newpage

\section{III  Second derivative of the conductance}

\begin{figure}[h!] \centering
\begin{minipage} [c] {0.25\textwidth}
\centering \includegraphics [width=2.5in, bb = 0 180 820 620 ]{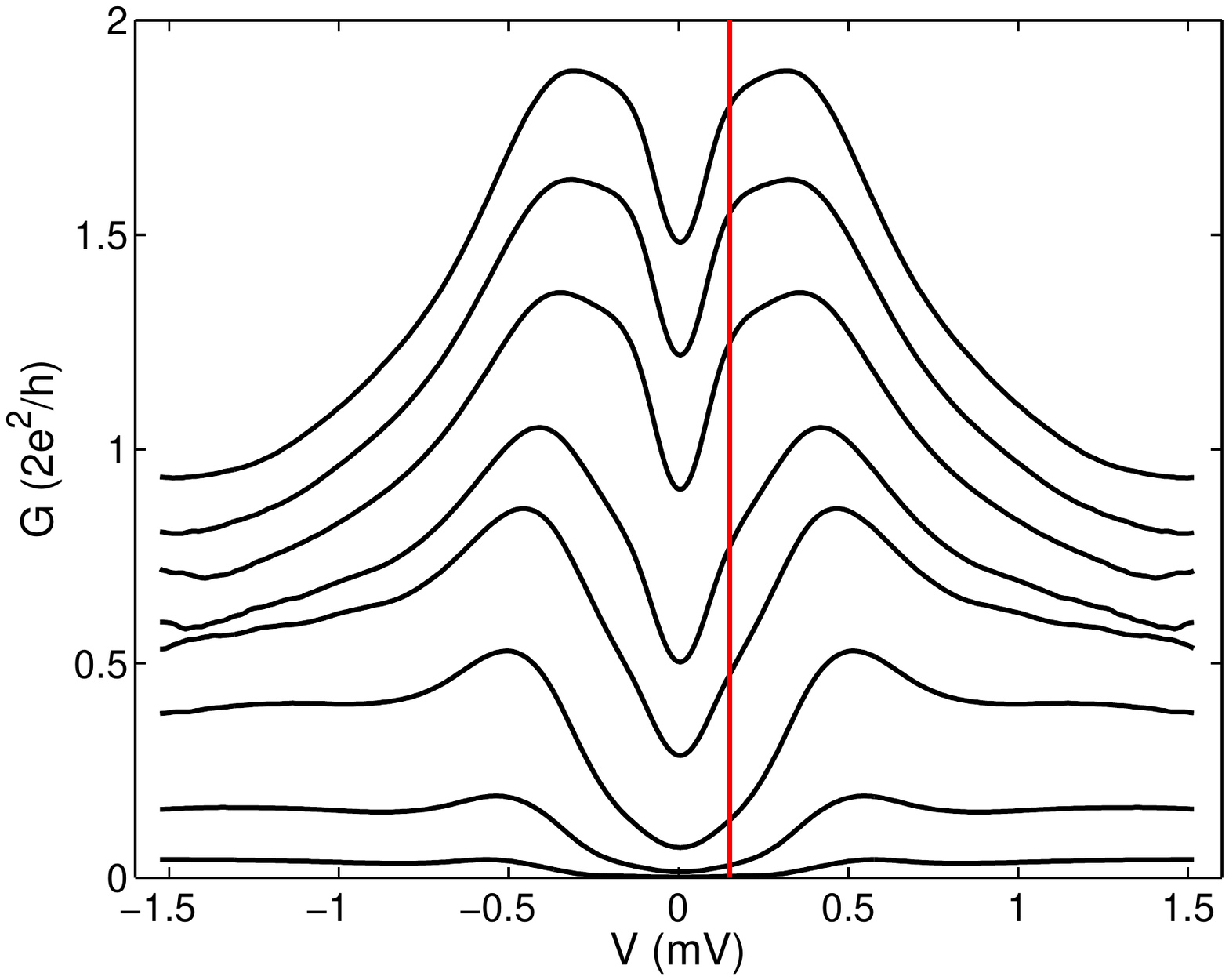}
\end{minipage}%
\begin{minipage} [c] {0.25\textwidth}
\centering \includegraphics [width=2.5in, bb = 0 180 820 620 ]{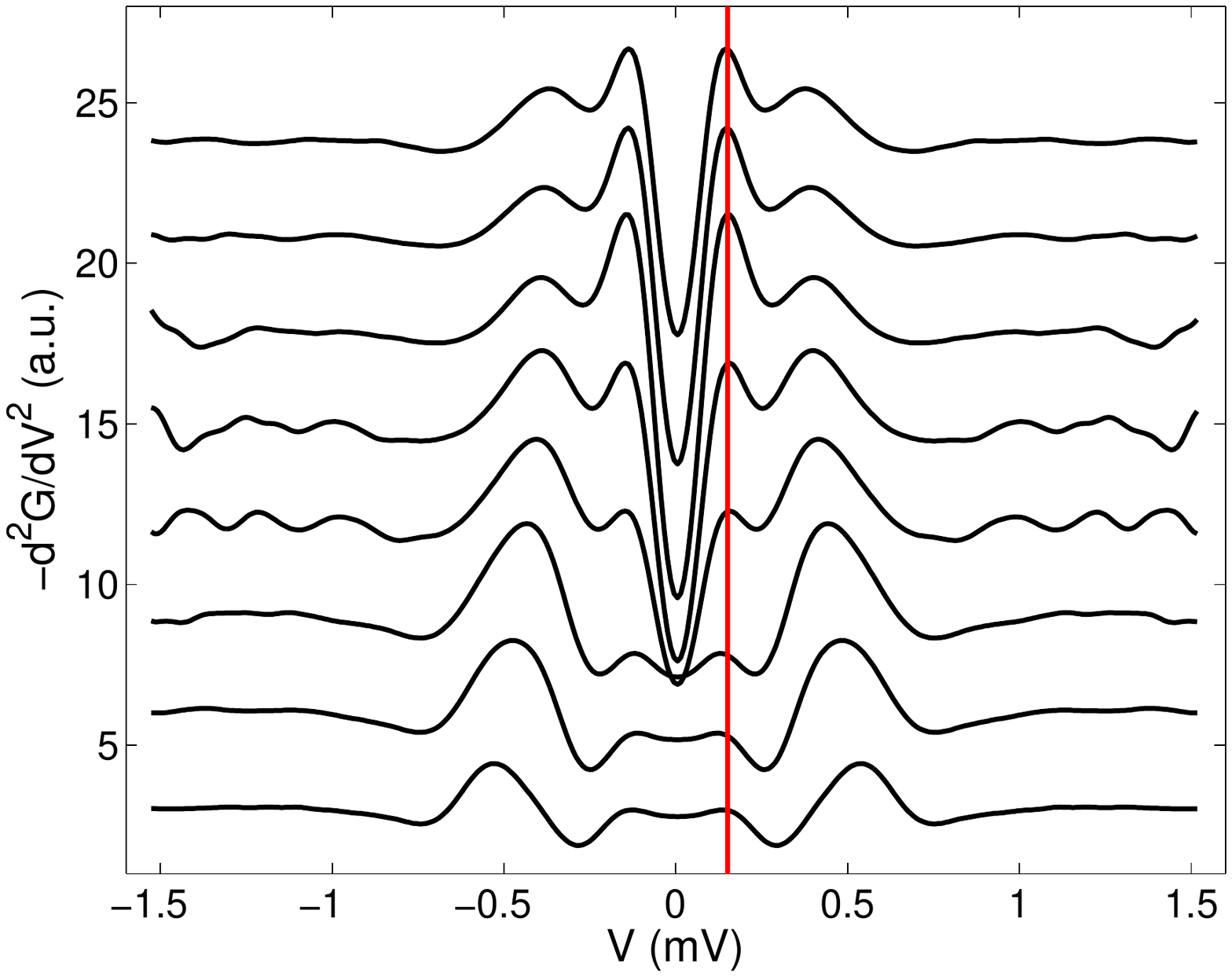}
\end{minipage}
\caption{{\bf Second derivative of an Andreev spectrum.} Left : Evolution of the raw local conductance $G(V)$ in $2e^2/h$ unit, for different values of the point-contact conductance acquired on InO-3 film. Right : second derivative $-d^2G/dV^2$ (arbitrary unit) of the same data. The vertical red lines point out the Andreev energy $\Delta_{c}$  in both panels.}
\label{figureS3}
\end{figure}

The energy, $\Delta_{c}$, can be readily evaluated with the position of the Andreev peaks in point-contact spectra $G(V)$ acquired at low conductance such as the one shown in the top curve of Figure 1 of the main text where these peaks are well defined. The energy of the Andreev peaks can be better resolved by inspecting the second derivative $-d^2G/dV^2$ as illustrated in Figure 2 of the main text. This proves also to be useful when the peaks at $E_{g}$ and $\Delta_{c}$ nearly overlap and are difficult to resolve. An example is shown in the point-contact evolution of Figure~\ref{figureS3}. The Andreev peaks are signaled by a change of slope in the raw conductance ( red vertical line in the left panel) which becomes more pronounced when the point-contact conductance increases. In the second derivative of the conductance, at this same energy a well-defined peak grows which enables a more accurate determination of $\Delta_{c}$.  

\section{IV Characterization of the low disordered film}

\begin{figure}[h!]
\includegraphics[width=1\linewidth, bb = 50 50 740 530 ]{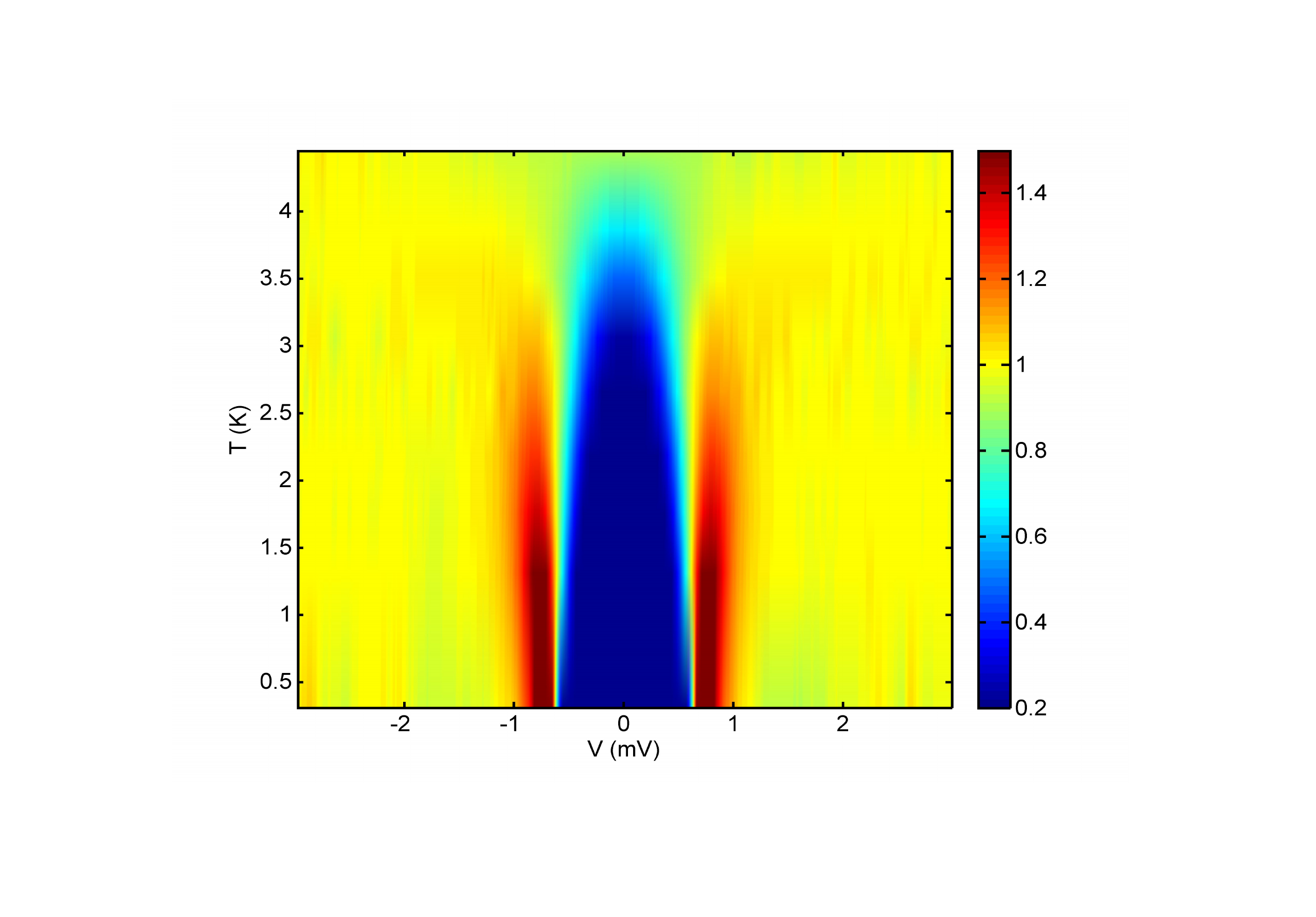}
\caption{{\bf Absence of preformed pairs in low disorder sample.} Colorplot of the T-evolution of the normalized tunneling conductance for InO-4 which shows the absence of pseudogap far above $T_c$ (3.2K). }
\label{figureS4}
\end{figure}

This section is devoted to a detailed characterization of InO-4 sample, whose disorder, according to its $T_c$ (see supplemental section I), is much weaker than the critical disorder of the superconductor-insulator transition in a:InO films. The thermal evolution of the one-particle density of states was measured in tunneling regime (tunneling conductance of $0.09 \,e^2/h$) and is displayed in the two-dimensional color plot of  Figure~\ref{figureS4}. A BCS fit (not shown) of the tunneling spectrum acquired at 50 mK gives a superconducting gap of $670 \mu eV $. This leads to a ratio $E_{g}/k_BT_c=2.4$ much smaller than the typical one measured in the three highly disordered samples. The superconducting gap closes slightly above $T_c$ in a temperature range which corresponds to the thermal fluctuation regime for disordered superconductors. There is therefore neither preformed Cooper pairs nor strong pseudogap regime far above $T_c$ in contrast with samples InO-1, InO-2 and InO-3 that are close to the superconductor-insulator transition~\cite{Sacepe11}.

\begin{figure}[h!]
\includegraphics[width=1\columnwidth, bb = 50 150 800 440 ]{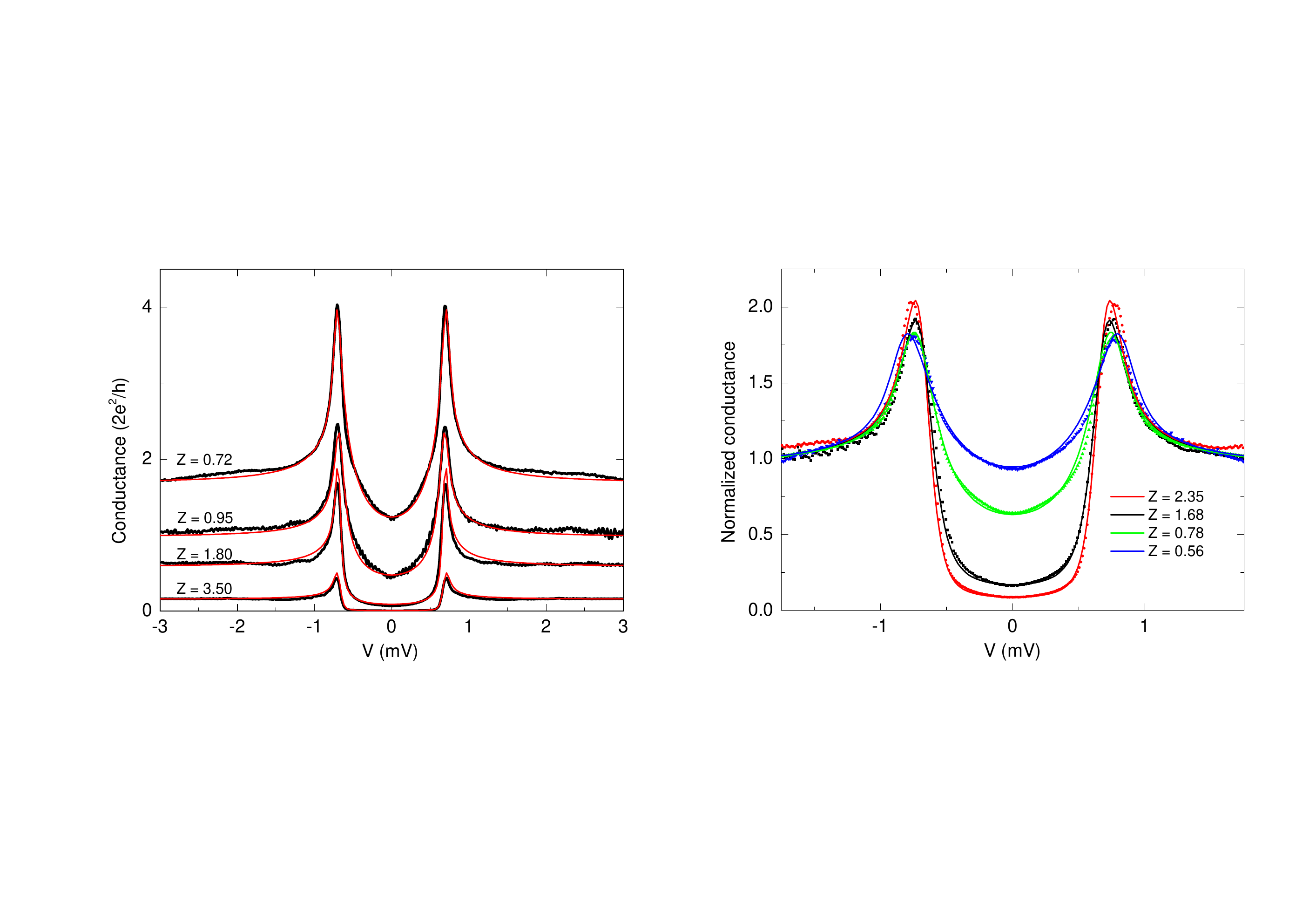}
\caption{{\bf Andreev Spectroscopy of a low disordered a:InO film}. The dotted curves are experimental data and the lines are BTK fits with a fitting parameter Z which represents the tunneling barrier height. Left : the conductance has been expressed in unit of the quantum of conductance for a comparison with Figure 1 of the main text. Right : another set of data normalized at high bias.}
\label{figureS5}
\end{figure}

We performed Andreev spectroscopy on this low disorder sample. Figure~\ref{figureS5} shows typical differential conductance curves from tunneling to Andreev regime. In the Andreev regime, the in-gap conductance increases continuously, without additional peaks. Those Andreev spectra can be accurately fitted with the Blonder-Tinkham-Klapwijk theory (BTK)~\cite{BTK82} which enables to extract the barrier strength Z. This conventional behavior proves that far from the superconductor-insulator transition, a single energy scale controls the transition to the superconductor as expected in BCS theory.

\end{document}